\documentclass[9pt,conference]{IEEEtran}

\usepackage{waspaa25} 

\usepackage{bm} 
\usepackage{acronym}
\usepackage{multirow}
\usepackage{amssymb} 

\usepackage{array}         
\usepackage{adjustbox}

\usepackage{xspace} 
\usepackage{threeparttable} 
\usepackage{tabularx}
\usepackage{enumitem}

\newcommand{\proposed}{VoxATtack\xspace}
\newcommand{\baselineOurs}{$\text{ECAPA}_{\text{ours}}$\xspace}
\newcommand{\baseline}{$\text{ECAPA}_{\text{baseline}}$\xspace}
\newcommand{\anonymization}{anonymization\xspace}
\newcommand{\anonymized}{anonymized\xspace}

\definecolor{pastelgreen}{RGB}{208,239,129}
\definecolor{lavender}{RGB}{222,210,234}
\definecolor{peach}{RGB}{255,198,142}
\newcommand{\graycell}[1]{\cellcolor{gray!20}#1}

\makeatletter
\long\def\@makefntext#1{\parindent 0pt\noindent\footnotesize#1}
\makeatother

\title{\proposed: A Multimodal Attack on Voice Anonymization Systems}

\name{Ahmad Aloradi$^{1,2}$,
      \"Unal Ege Gaznepoglu$^{2}$,
      Emanu\"el A.~P.~Habets$^{2}$,
      Daniel Tenbrinck$^{1}$\thanks{This work was funded by the German Ministry of Research, Technology and Space (BMFTR) under grant agreement No. 16IS24072F (COMFORT).}}
\address{
$^{1}$Department of Data Science; $^{2}$International Audio Laboratories Erlangen, FAU Erlangen-N\"urnberg, Germany
}

\begin{document}
\bstctlcite{bstctl}  

\newacro{ICASSP}{International Conference on Acoustics, Speech, and Signal Processing}
\newacro{AAM}{Additive Angular Margin Softmax}
\newacro{AS-norm}{adaptive score normalization}
\newacro{LoRA}{low-rank adaptation}
\newacro{PLDA}{probabilistic linear discriminant analysis}
\newacro{BN}{bottleneck feature}
\newacro{RTF}{real time factor}
\newacro{DNN}{deep neural network}
\newacro{F0}{fundamental frequency}
\newacro{EER}{equal error rate}
\newacro{WER}{word error rate}
\newacro{PI}{personal information}
\newacro{MSE}{mean-squared error}
\newacro{NSF}{neural source-filter}
\newacro{AM}{acoustic model}
\newacro{VPC}{VoicePrivacy Challenge}
\newacro{VPAC}{VoicePrivacy Attacker Challenge}
\newacro{SV}{speaker verification}
\newacro{ASV}{automatic speaker verification}
\newacro{ASR}{automatic speech recognition}
\newacro{TDNN}{time delay neural network}
\newacro{FPE}{Fine Pitch Error}
\newacro{GPE}{Gross Pitch Error}
\newacro{MFCC}{Mel Frequency Cepstral Coefficients}
\newacro{RTF}{Real Time Framework}
\newacro{NSF}{Neural Source-Filter}
\newacro{VC}{voice conversion}
\newacro{AE}{autoencoder}
\newacro{CNN}{convolutional neural network}
\newacro{PPG}{phonetic posteriorgrams}
\newacro{UAR}{unweighted average recall}
\newacro{PQMF}{pseudo quadrature mirror filterbank}
\newacro{OOD}{out-of-distribution}
\newacro{SER}{speech emotion recognition}
\newacro{OHNN}{orthogonal Householder neural network}
\newacro{GAN}{generative adversarial network}
\newacro{AAM}{additive angular margin}
\newacro{ZEBRA}{Zero Evidence Biometric Recognition Assessment}

\maketitle

\begin{abstract}
Voice anonymization systems aim to protect speaker privacy by obscuring vocal traits while preserving the linguistic content relevant for downstream applications. However, because these linguistic cues remain intact, they can be exploited to identify semantic speech patterns associated with specific speakers.
In this work, we present \proposed, a novel multimodal de-anonymization model that incorporates both acoustic and textual information to attack anonymization systems.
While previous research has focused on refining speaker representations extracted from speech, we show that incorporating textual information with a standard ECAPA-TDNN improves the attacker's performance.
Our proposed \proposed model employs a dual‑branch architecture, with an ECAPA‑TDNN processing anonymized speech and a pretrained BERT encoding the transcriptions.
Both outputs are projected into embeddings of equal dimensionality and then fused based on confidence weights computed on a per-utterance basis.
When evaluating our approach on the \ac{VPAC} dataset, it outperforms the top-ranking attackers on five out of seven benchmarks, namely B3, B4, B5, T8-5, and T12-5. To further boost performance, we leverage anonymized speech and SpecAugment as augmentation techniques. This enhancement enables \proposed to achieve state-of-the-art on all \ac{VPAC} benchmarks, after scoring 20.6\% and 27.2\% average \acl{EER} on T10-2 and T25-1, respectively.
Our results demonstrate that incorporating textual information and selective data augmentation reveals critical vulnerabilities in current voice anonymization methods and exposes potential weaknesses in the datasets used to evaluate them.
\end{abstract}

\section{Introduction}
\label{sec:intro}

Speech can reveal sensitive personal attributes, including identity, age, gender, ethnicity, and health, through both human perception and automated systems \cite{nautsch_preserving_2019, tomashenko2024first}.
The potential for information disclosure raises critical privacy challenges that intersect with regulatory frameworks, e.g., the European General Data Protection Regulation (GDPR).

In response to privacy risks, voice \anonymization (or simply \anonymization) systems have been developed to obscure the speaker's identity while preserving information for downstream tasks such as \ac{ASR} \cite{champion_vpc_evalplan_2024}. 
The VoicePrivacy initiative, launched in 2020, organized three \acp{VPC} to foster the development of anonymization systems \cite{tomashenko20_interspeech, tomashenko_vpc_evalplan_2022, champion_vpc_evalplan_2024}.
In 2025, the first VoicePrivacy \emph{Attacker} Challenge (VPAC) shifted focus to the evaluation of state-of-the-art anonymization systems \cite{tomashenko2024first}.

The aim of the \acs{VPAC} is to develop attacker systems capable of undermining the privacy protections of \anonymization frameworks. Specifically, participants are tasked with designing an \ac{ASV} model which, given two \anonymized speech utterances, determines whether they originate from the same speaker.
Participants were instructed to develop their attacker systems against one or more \anonymization systems chosen from three \ac{VPC} 2024 baselines (B3, B4, B5) and four other top-performing participant-submitted systems (T8-5, T10-2, T12-5, T25-1) \cite{tomashenko2024first}.

The challenge baseline is an attacker based on ECAPA-TDNN \cite{desplanques_ecapa-tdnn_2020} and trained by the participants on data \anonymized using their \anonymization system, i.e., in a semi-informed setting. The results in \cite{tomashenko2024first} suggest that the baseline struggles with the \anonymized samples, scoring worse than $40\%$ \ac{EER} against all the top-performing \anonymization systems ($50\%$ \ac{EER} is random chance).

Recently, the \acs{VPAC} results have been announced, revealing the top-five performing attackers \cite{best_baseline, zhang_attacking_2025, HLTCOE, SpecWav, fine_tune_titanet}. Among these, A.20 was the top-ranking attacker against all systems except T8-5 \cite{best_baseline}. The authors achieved this via a multistage training strategy to fine-tune ResNet34 on \ac{ASV} using \acl{LoRA}, while enhancing its output by incorporating WavLM \cite{chen2022wavlm} features through the residual connections.
Against T8-5, attacker A.5 \cite{zhang_attacking_2025} achieved the best results using ECAPA-TDNN and \ac{PLDA} scoring, with contrastive loss regularization and data augmentation via SpecAugment \cite{park19e_interspeech} and the original (non-\anonymized) speech.
Similarly, \cite{HLTCOE} evaluated ECAPA-TDNN with multiple scoring metrics, e.g., \ac{PLDA} and maximum similarity. However, they additionally applied voice normalization via kNN-VC \cite{baas2023voice} when attacking system T10-2.
In \cite{SpecWav}, ECAPA-TDNN was also adopted and trained in two stages, with the feature extractor replaced by wav2vec~$2.0$ outputs \cite{wav2vec2} and spectrogram resizing applied for data augmentation.
In \cite{fine_tune_titanet}, TitaNet-Large \cite{koluguri2022titanet} and ECAPA-TDNN were investigated under various training data conditions. Crucially, their results demonstrated that finetuning TitaNet-Large on all \anonymized data simultaneously proved ineffective, suggesting the importance of considering the specific \anonymization system when training an attacker.
Finally, the work in \cite{10887896}, although not part of the challenge, showed the effectiveness of using phoneme durations for speaker de-\anonymization.

In this work, we investigate different techniques to improve attacks against \anonymization systems. In particular, we study the effects of incorporating textual content and data augmentation to enhance de-\anonymization.
We propose a multimodal approach that leverages audio and textual information for the attacks. We hereafter refer to our attacker model as \emph{\proposed} (short for \emph{Voice and Text Attack}).

The motivation to investigate text as an auxiliary modality was inspired by our recent work in \cite{ege2025}, which showed that, on LibriSpeech \cite{panayotov_librispeech_2015}, text alone can be indicative of speaker identities due to topic similarities\footnote{While in \cite{ege2025} we demonstrated that topic similarity in LibriSpeech enables distinguishing speakers, we emphasize that in general \textit{transcribed speech} contains other potentially identifying features such as filler words (e.g., "um") and discourse markers (e.g., "well, I mean") \cite{speech-aa2023}. That said, text does not always convey speaker-identifying information, e.g., in text-dependent \ac{ASV}.}.
Our experiments demonstrate that our multimodal approach outperforms the top-performing systems in the \ac{VPAC} on B3, B4, B5, T12-5, and T8-5.
When further refining our training by applying data augmentation with \anonymized speech and SpecAugment \cite{park19e_interspeech}, our \proposed model also outperforms the \ac{VPAC} winner on the remaining two systems: T10-2 and T25-1. 
Our code and experimental results are publicly available\footnote{\url{https://github.com/ahmad-aloradi/VoxATtack}.}.

\section{Proposed Attacker Model}

\setlength{\textfloatsep}{0pt} 
\setlength{\intextsep}{0pt}    

\begin{figure}[t]
    \centering
    \adjustbox{center,  padding*=6ex 0ex 0ex 0ex}{\includegraphics[width=1.6\columnwidth]{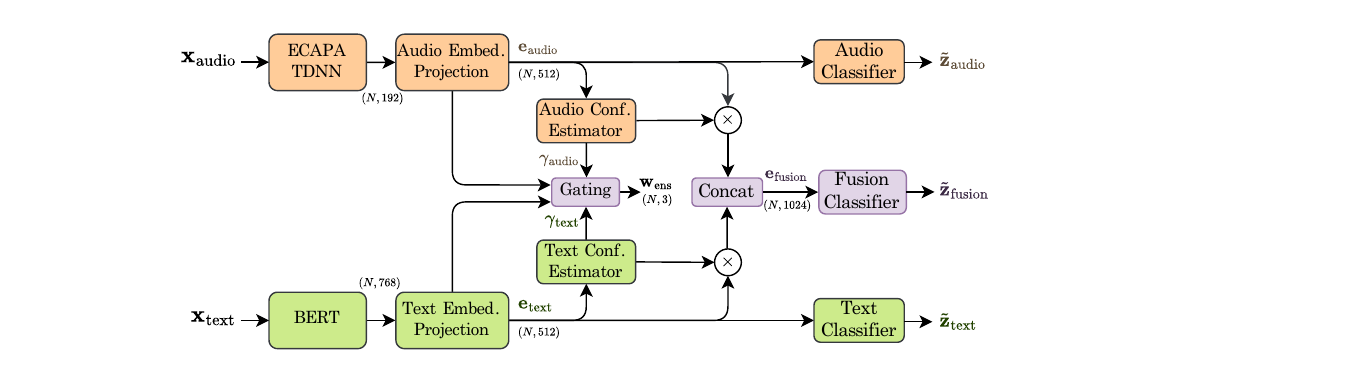}}
    \caption{Architecture of the proposed \proposed. $\mathbf{x}_{\text{audio}}$ and $\mathbf{x}_{\text{text}}$ refer to the audio and text pairs. $\mathbf{w}_{\text{ens}}$ is used to combine the predictions of the three classifiers. During inference, $\mathbf{e}_{\text{fusion}}$ is used as an embedding}
    \label{fig:model}
\end{figure}

\begin{table}[t]
\centering
\caption{Details of the proposed \proposed architecture}
\label{tab:model}
\Large
\resizebox{\columnwidth}{!}{%
\begin{tabular}{llr}
\toprule
\textbf{Component} & \textbf{Architecture} & \textbf{Parameters} \\
\midrule
\rowcolor{peach}
\multicolumn{3}{l}{\textbf{Audio Branch}} \\
\midrule
Audio Embedding Model  & ECAPA-TDNN \cite{desplanques_ecapa-tdnn_2020} (Audio $\rightarrow$ $192$) & (fine-tune) $22.2$~M \\
Audio Embedding Projection    & LayerNorm, FC ($192$ $\rightarrow$ $512$), GELU, Dropout ($p=0.3$) & $0.1$~M \\
Audio Confidence Estimator   & FC ($512$ $\rightarrow$ $256$), ReLU, FC ($256$ $\rightarrow$ $1$), Sigmoid & $0.1$~M \\
\midrule
\rowcolor{pastelgreen}
\multicolumn{3}{l}{\textbf{Text Branch}} \\
\midrule
Text Embedding Model   & BERT \cite{devlin_bert_2019} (Text $\rightarrow$ $768$) & (fixed) $109.0$~M \\
Text Embedding Projection   & LayerNorm, FC ($768$ $\rightarrow$ $512$), GELU, Dropout ($p=0.1$) & $0.4$~M \\
Text Confidence Estimator   & FC ($512$ $\rightarrow$ $256$), ReLU, FC ($256$ $\rightarrow$ $1$), Sigmoid & $0.1$~M \\
\midrule
\rowcolor{lavender}
\multicolumn{3}{l}{\textbf{Fusion \& Ensemble Prediction}} \\
\midrule
Embeddings Fusion      & Concat($\gamma_{\text{audio}} \times \mathbf{e}_{\text{audio}}$, $\gamma_{\text{text}} \times \mathbf{e}_{\text{text}}$) & - \\
Gating                 & FC ($1026$ $\rightarrow$ $512$), ReLU, FC ($512$ $\rightarrow$ $3$), Softmax & $0.5$~M \\
\midrule
Classifiers            & FC ($512$/$512$/$1024$ $\rightarrow$ $1001$), Softmax & $2.1$~M \\
\midrule
\midrule
\multicolumn{2}{l}{\textbf{Trainable Parameters}} &  $\mathbf{25.5}$~\textbf{M} \\
\bottomrule
\end{tabular}
} 
\end{table}

In the following, we introduce \proposed model for speaker de-\anonymization. The model leverages both acoustic and textual modalities simultaneously to identify speakers from \anonymized speech paired with corresponding transcriptions.
While standard \ac{ASV} systems primarily rely on acoustic signals that naturally contain speaker-specific characteristics, \anonymization techniques deliberately obscure these cues. 
Our approach attempts to compensate for this by also exploiting textual content. Such content may contain distinctive idiolectal patterns and personal linguistic habits that remain unchanged when voice characteristics are disguised.

\subsection{Model architecture}
We adopt a straightforward architectural design to highlight the effectiveness of incorporating text as an auxiliary input. Moreover, we rely on ECAPA-TDNN for extracting speech embeddings to remain consistent with the baseline system.
The overall architecture of our \proposed model is illustrated in \Cref{fig:model}, with detailed specifications in \cref{tab:model}. Our model can be divided into three main components:
\begin{itemize}[leftmargin=8.5pt]
    \item \textbf{Audio branch:} This branch utilizes a pretrained ECAPA-TDNN to extract speaker representations from \anonymized speech. It then projects these representations into $512$-dimensional embeddings $\mathbf{e}_{\text{audio}}$ via a projection layer. The branch includes a confidence estimator that produces a reliability score for $\mathbf{e}_{\text{audio}}$, denoted as $\gamma_{\text{audio}}$. A sigmoid activation is applied to the confidence values, ensuring their range is normalized to the interval $(0,\,1)$.

    \item \textbf{Text branch:} This branch processes the linguistic content of utterances, leveraging a pretrained BERT model to extract contextual representations from transcriptions. BERT's contextualized token embeddings are pooled to obtain a sentence-level representation, which captures the lexical choices and linguistic patterns characteristic of individual speakers. This representation is then projected through a text projection layer into a $512$-dimensional embedding $\mathbf{e}_{\text{text}}$. Similar to the audio branch, a confidence estimator produces a score $\gamma_{\text{text}}$ based on $\mathbf{e}_{\text{text}}$.

    \item \textbf{Fusion:} To fuse information from both modalities, we apply confidence-based weighting to each embedding. Specifically, the weighted embeddings are computed as $\gamma_{\text{audio}} \times \mathbf{e}_{\text{audio}}$ and $\gamma_{\text{text}} \times \mathbf{e}_{\text{text}}$, respectively. These weighted representations are then concatenated to form a fused embedding $\mathbf{e}_{\text{fusion}} \in \mathbb{R}^{1024}$. This mechanism allows the model to dynamically adjust the contribution of each modality based on its estimated reliability.
\end{itemize}

Each branch has a dedicated classifier to encourage its respective embedding to be discriminative, which helps when evaluating modalities individually.
During training, an ensemble prediction is obtained by computing a weighted sum of the classifier outputs (logits). Specifically, the ensemble prediction $\tilde{\mathbf{z}}_{\text{ens}}$ is computed as follows:
\begin{equation}
\tilde{\mathbf{z}}_{\text{ens}} = w_f\tilde{\mathbf{z}}_{\text{fusion}} + w_a\tilde{\mathbf{z}}_{\text{audio}} + w_t\tilde{\mathbf{z}}_{\text{text}},
\end{equation}
$\mathbf{w}_{\text{ens}} = [w_f, w_a, w_t] \in \mathbb{R}^{3}$ are the ensemble weights, with $w_i \in [0,1]$ and $\sum_i w_i = 1$, and $\tilde{\mathbf{z}} \in \mathbb{R}^{C}$ denotes each classifier's output, with $C$ denoting the number of training speakers.
The ensemble weights are produced by a gating mechanism that takes as input the concatenation of $\mathbf{e}_{\text{audio}}$ and $\mathbf{e}_{\text{text}}$ along with their confidence scores $\gamma_{\text{audio}}$ and $\gamma_{\text{text}}$.
This way, the model dynamically adapts the contribution of each modality to improve accuracy while avoiding overreliance on a single source of information.

\subsection{Training}
We fine-tune the ECAPA-TDNN parameters to adapt to \anonymization distortions while keeping the BERT encoder weights fixed to preserve its pretrained linguistic knowledge. The rest of the model is trained from scratch. 
Our model is trained by optimizing a multi-task objective composed of four \ac{AAM} loss terms \cite{xiang_margin_2019}, one for each of the audio, text, and fusion classifiers, plus one for the ensemble output. The training objective is formulated as:
\begin{equation}
    \label{eqn:our_loss}
    L = \lambda_e L_{\text{ensemble}} + \lambda_f L_{\text{fusion}} + \lambda_a L_{\text{audio}} + \lambda_t L_{\text{text}},
\end{equation}
where $\lambda_e$, $\lambda_f$, $\lambda_a$, and $\lambda_t$ control the contribution of the associated \ac{AAM} loss term.

\subsection{Inference}
When using our model at inference, we discard the classification heads and use the fused embedding $\mathbf{e}_{\text{fusion}}$ as speaker representation. For enrollment, we compute a centroid for each enrolled speaker by averaging the embeddings extracted from all their enrollment utterances. Embeddings are $\text{L}2$-normalized before computing the cosine similarity scores. Speaker verification is then performed by calculating the cosine similarity between the trial embedding and each enrolled speaker's centroid.

\section{Experimental Setup}
\label{sec:exp_setup}

\subsection{Dataset and evaluation metrics}
\label{subsec:data}
We use the \ac{VPAC} corpus, which consists of LibriSpeech data \cite{panayotov_librispeech_2015} \anonymized by seven systems (B3, B4, B5, T8-5, T10-2, T12-5, T25-1).
The dataset provides transcriptions corresponding to LibriSpeech. We use these as ground-truth text labels, since (1) the anonymization systems accurately preserve linguistic content by design \cite{voiceprivacy2024slides}, and (2) \ac{ASR} would conflate anonymization quality with intelligibility.

Both the development and test sets are gender-segregated and split into \ac{ASV} trials, each with a similar distribution: Approximately $95.4\%$ negative trials (impostor vs.\ target) and $4.6\%$ positive trials (target vs.\ target).
We follow the \ac{VPAC} evaluation protocol and report the \acp{EER} on the development and test sets, which represent the mean \acp{EER} across male and female trials. We also report $\textrm{EER}_{\textrm{avg}}$, which is the average \ac{EER} across the development and test sets.
 
\subsection{Preprocessing}
\label{subsec:preprocessing}
Speech segments were cropped to a maximum duration of $10$ seconds during training.
The resulting segments were mean-centered and normalized to $-20$~dB root mean square level, with maximum amplitude clipping at $1.0$.
The corresponding transcriptions were left untruncated to preserve full contextual information.
Although this introduces a mismatch in the duration of audio and text inputs, our architecture handles this gracefully by processing and aggregating each modality independently prior to fusion.

\subsection{Implementation details}
\label{subsec:setup_model}
For audio processing, we extract $192$-dimensional speaker embeddings using SpeechBrain’s ECAPA-TDNN implementation\footnote{\url{https://huggingface.co/speechbrain/spkrec-ecapa-voxceleb}.} \cite{speechbrain}, which is pretrained on VoxCeleb~1 and~2 \cite{Chung18b}.
For text analysis, we employ the pretrained BERT-base model from Hugging Face\footnote{\url{https://huggingface.co/google-bert/bert-base-uncased}.} \cite{devlin_bert_2019}. To obtain a contextualized sentence-level embedding, we use the $768$-dimensional pooled representation from the \texttt{[CLS]} token output.
Sentence-level text embeddings offer the additional benefit of robustness in scenarios where only \ac{ASR} transcriptions are available.
Both audio and text embeddings are projected into a $512$-dimensional space before fusion.

The extracted embeddings are $\text{L}2$-normalized before computing the similarity scores. Cosine similarities are also normalized using \ac{AS-norm} \cite{matejka17_interspeech}, with the top $1000$ cohort embeddings drawn from $40$k random training samples. 
For system T8-5, however, we skip \ac{AS-norm} since it randomly applies one of two different \anonymization methods, rendering consistent cohort matching ineffective in this case.

In our loss function \cref{eqn:our_loss}, we use equal weighting for the auxiliary terms $\lambda_f = \lambda_a = \lambda_t = 0.1$, and $\lambda_e = 1$ for the primary ensemble loss. For the \ac{AAM} parameters, we set the scale and margin to $30$ and $0.15$, respectively. We employed the AdamW optimizer \cite{loshchilov2018decoupled} with a weight decay factor of $10^{-5}$. \texttt{CyclicLR} \cite{cyclic_lr} was used to cycle the learning rate at every optimization step between $10^{-4}$ and $10^{-6}$, completing a full cycle every $13$k steps. The batch size was set to $32$ and the maximum number of training epochs was $25$, with an early stopping after $10$ consecutive epochs without \ac{EER} reduction on the validation set.

To isolate the impact of including text, we train our audio branch alone, excluding both the text and fusion modules. 
Since the resulting architecture is similar, though not identical, to the official baseline model \baseline, we refer to our audio baseline as \baselineOurs. The training settings for \baselineOurs remain the same as those used for \proposed, 
except that we optimize one \ac{AAM} loss term ($L_{\text{audio}}$).

For the augmentations in \cref{sec:augmentation}, we used time and frequency masking together with a $\pm10\%$ speed perturbation \cite{speechbrain}. 
Each augmented utterance generates a new sample for each of the three transformations. To maintain the same batch size, we load $8$ original samples from the \anonymized datasets and generate the remaining samples using the three augmentations applied to each original sample.

\section{Results}
\label{sec:results}


\begin{table}[t]
\centering
\caption{Performance of several attacker systems in $\textrm{EER}_{\textrm{avg}}$~$[\%]$~\mbox{$(\downarrow)$}. \baseline refers to the official baseline, while \baselineOurs is our audio-only system. Systems A.5 and A.20 together achieve the top scores against every \anonymization system in the \ac{VPAC} data}
\label{tab:performance_comparison}
\resizebox{0.9\columnwidth}{!}{
\begin{threeparttable}
\renewcommand{\arraystretch}{0.77}
\setlength{\tabcolsep}{4pt}
\footnotesize
\sisetup{
reset-text-series  = false,
text-series-to-math= true,
mode               = text,
tight-spacing      = true,
round-mode         = places,
round-precision    = 1,
table-format       = 2.2,
table-number-alignment = center,
detect-weight=true,
}

\begin{tabular*}{\columnwidth}{@{\extracolsep{\fill}}
  c
  l 
  S[table-format=2.2]
  S[table-format=2.2]
  S[table-format=2.2]
}
\toprule
\multirow{2}{*}{Anon. dataset} & \multirow{2}{*}{Attacker system} & \multicolumn{3}{c}{\ac{EER}~$[\%]$~$(\downarrow)$} \\
\cmidrule(lr){3-5}
 & & {Dev.} & {Test} & {Average} \\
\midrule
\multirow{4}{*}{B3} 
  & \baseline  \cite{champion_vpc_evalplan_2024} & 25.24 & 27.32 & \graycell{26.28} \\
  & A.5 \cite{zhang_attacking_2025}        & 23.55 & 24.47 & \graycell{24.01} \\
  & A.20 \cite{best_baseline}       & 23.70 & 20.50 & \graycell{22.10} \\
  & \baselineOurs  & 22.27 & 21.39 & \graycell{21.83} \\
  & \proposed      & \bfseries 21.72 & \bfseries 19.91 & \graycell{\bfseries 20.81} \\
\midrule
\multirow{4}{*}{B4} 
  & \baseline \cite{champion_vpc_evalplan_2024} & 32.71 & 30.26 & \graycell{31.49} \\
  & A.5 \cite{zhang_attacking_2025}        & 25.49 & 21.26 & \graycell{23.38} \\
  & A.20 \cite{best_baseline}       & \bfseries 22.10 & 19.50 & \graycell{20.80} \\
  & \baselineOurs  & 23.24 & 19.25 & \graycell{21.25} \\
  & \proposed      & 22.82 & \bfseries 18.35 & \graycell{\bfseries 20.58} \\
\midrule
\multirow{4}{*}{B5} 
  & \baseline \cite{champion_vpc_evalplan_2024} & 34.37 & 34.34 & \graycell{34.36} \\
  & A.5 \cite{zhang_attacking_2025}        & 30.16 & 27.48 & \graycell{28.82} \\
  & A.20 \cite{best_baseline}       & \bfseries 27.56 & 25.45 & \graycell{26.55} \\
  & \baselineOurs  & 30.98 & 26.26 & \graycell{28.62} \\
  & \proposed      & 27.82 & \bfseries 24.31 & \graycell{\bfseries 26.09} \\
\midrule
\multirow{4}{*}{T10-2} 
  & \baseline \cite{champion_vpc_evalplan_2024} & 41.83 & 40.63 & \graycell{41.84} \\
  & A.5 \cite{zhang_attacking_2025}\tnote{1} & {*} & {*} & \graycell{32.23} \\
  & A.20 \cite{best_baseline}       & \bfseries 23.75 & \bfseries 22.55 & \graycell{\bfseries 23.20} \\
  & \baselineOurs  & 30.55 & 30.27 & \graycell{30.41} \\
  & \proposed      & 29.25 & 28.74 & \graycell{28.99} \\
\midrule
\multirow{4}{*}{T12-5}
  & \baseline \cite{champion_vpc_evalplan_2024} & 43.71 & 42.75 & \graycell{43.23} \\
  & A.5 \cite{zhang_attacking_2025}        & 30.78 & 27.14 & \graycell{28.96} \\
  & A.20 \cite{best_baseline}       & 28.75 & 25.50 & \graycell{27.13} \\
  & \baselineOurs  & 31.43 & 26.79 & \graycell{29.11} \\
  & \proposed      & \bfseries 28.37 &  \bfseries 25.14 & \graycell{\bfseries 26.75} \\
\midrule
\multirow{4}{*}{T25-1}
  & \baseline \cite{champion_vpc_evalplan_2024} & 41.36 & 42.13 & \graycell{41.75} \\
  & A.5 \cite{zhang_attacking_2025}        & 33.31 & 32.82 & \graycell{33.07} \\
  & A.20 \cite{best_baseline}       & \bfseries 29.05 & \bfseries 27.70 & \graycell{\bfseries 28.40} \\
  & \baselineOurs  & 33.34 & 31.41 & \graycell{32.38} \\
  & \proposed      & 31.89 &  30.28 & \graycell{31.08} \\
\midrule
\multirow{4}{*}{T8-5}
  & \baseline \cite{champion_vpc_evalplan_2024} & 40.24 & 41.28 & \graycell{40.76} \\
  & A.5 \cite{zhang_attacking_2025}        & 27.24 & 24.85 & \graycell{26.05} \\
  & A.20 \cite{best_baseline}       & 29.10 & 26.30 & \graycell{27.70} \\
  & \baselineOurs  & 27.31 & 27.42 & \graycell{27.37} \\
  & \proposed      & \bfseries 23.66 & \bfseries 24.50 & \graycell{\bfseries 24.08} \\
\bottomrule
\end{tabular*}
\begin{tablenotes}
    \footnotesize
    \item[1] Only $\textrm{EER}_{\textrm{avg}}$ was reported for T10-2. Authors also replaced their attacker by TitaNet-Large with cosine similarity scoring
\end{tablenotes}
\end{threeparttable}
}
\end{table}

\subsection{Attacker systems evaluation}

\label{sec:main_results}
The performance of the attacker systems is reported in \cref{tab:performance_comparison}. 
Our \proposed model consistently outperformed other attacker models, demonstrating the best performance on five of the seven evaluated benchmarks (B3, B4, B5, T12-5, and T8-5). 
The lowest overall \acp{EER} were observed on B3 and B4, achieving $20.8\%$ and $20.6\%$, respectively. The highest \acp{EER} were $31.1\%$ and $29.0\%$, which were obtained against T25-1 and T10-2, respectively. 
Compared to the \ac{VPAC} top-ranking systems, our attack model showed the best performance against all \anonymization systems except for T25-1 and T10-2, where A.20 scores the lowest \acp{EER}, with a particularly wide margin on T10-2.

Comparing \baseline and \baselineOurs revealed surprising results: Despite sharing the same ECAPA-TDNN backbone, performance gaps were substantial across all systems. \baselineOurs showed improvements ranging from $4.5\%$ \ac{EER} reduction on B3 to $14.1\%$ on T12-5.
This gap highlights the critical impact of hyperparameter choices on attacker performance, suggesting that careful tuning may be as important as architectural changes when evaluating \anonymization systems.

The consistent improvements from \baselineOurs to \proposed demonstrate the effectiveness of our multimodal approach.
The incorporation of textual information reduces \ac{EER} by $0.7\%$ on B4 and up to $3.3\%$ on T8-5.
These findings suggest exploitable intra-speaker linguistic patterns in the dataset, potentially exposing a vulnerability when using it in \anonymization applications.
To further validate our hypothesis, we fed only the transcriptions into the text branch and used $\mathbf{e}_{\text{text}}$ as speaker representation instead of $\mathbf{e}_{\text{fusion}}$. The resulting $\textrm{EER}_{\textrm{avg}}$ was $35.8\%$, which remarkably surpassed \baseline on all systems except B3, B4 and B5, despite the baseline having access to the \anonymized speech signals.


\begin{table}[t!]
\centering
\caption{\proposed performance against all anonymization systems and LibriSpeech in \ac{EER} $[\%]$~$(\downarrow)$.  The last two rows show \baselineOurs (trained on all systems) and \baseline (trained on LibriSpeech only)}
\label{tab:gen_attacks}
\renewcommand{\arraystretch}{0.8}
\scriptsize
\resizebox{\columnwidth}{!}{
\setlength{\tabcolsep}{1.0pt}
\sisetup{
reset-text-series  = false,
text-series-to-math= true,
mode               = text,
tight-spacing      = true,
round-mode         = places,
round-precision    = 1,
table-format       = 2.2,
table-number-alignment = center,
detect-weight=true,
}
\begin{tabular}{
  l
  l
  S[table-format=2.2]         
  S[table-format=2.2]         
  S[table-format=2.2]         
  S[table-format=2.2]         
  S[table-format=2.2]         
  S[table-format=2.2]         
  S[table-format=2.2]         
}
\toprule
\multirow{3}{*}{Attacker} & \multirow{3}{*}{Eval. data} & \multicolumn{3}{c}{Dev. set} & \multicolumn{3}{c}{Test set} & \multicolumn{1}{c}{\multirow{3}{*}{$\textrm{EER}_{\textrm{avg}}$}} \\
\cmidrule(lr){3-5} \cmidrule(lr){6-8}  &  & \multicolumn{1}{c}{Female} & \multicolumn{1}{c}{Male} & \multicolumn{1}{c}{Average} & \multicolumn{1}{c}{Female} & \multicolumn{1}{c}{Male} & \multicolumn{1}{c}{Average} &  \\
\midrule
\multirow{7}{*}{\proposed} & B3 & 26.17 & 19.73 & 22.95 & 25.18 & 14.92 & 20.05 & \graycell{21.50} \\
 & B4 & 25.77 & 23.73 & 24.75 & 20.44 & 17.81 & 19.13 & \graycell{21.94} \\
 & B5 & 29.47 & 26.69 & 28.08 & 25.55 & 22.49 & 24.02 & \graycell{26.05} \\
 & T10-2 & 24.22 & 21.80 & 23.01 & 21.40 & 18.26 & 19.83 & \graycell{21.42} \\
 & T12-5 & 29.35 & 28.03 & 28.69 & 27.57 & 21.85 & 24.71 & \graycell{26.70} \\
 & T25-1 & 31.32 & 29.09 & 30.21 & 30.40 & 24.62 & 27.51 & \graycell{28.86} \\
 & T8-5 & 27.02 & 24.97 & 26.00 & 25.30 & 21.06 & 23.18 & \graycell{24.59} \\
 & LibriSpeech & 7.22 & 2.64 & 4.93 & 4.56 & 2.22 & 3.39 & \graycell{4.16} \\
\midrule
\baselineOurs & LibriSpeech & 5.55 & 1.40 & 3.47 & 6.38 & 2.00 & 4.20 & \graycell{3.83} \\
\baseline \cite{champion_vpc_evalplan_2024} & LibriSpeech & 10.51 & 0.93 & 5.72 & 8.76 & 0.42 & 4.59 & \graycell{5.16} \\
\bottomrule
\end{tabular}
}
\end{table}

\begin{figure}
    \centering
    \includegraphics[width=0.9\columnwidth]{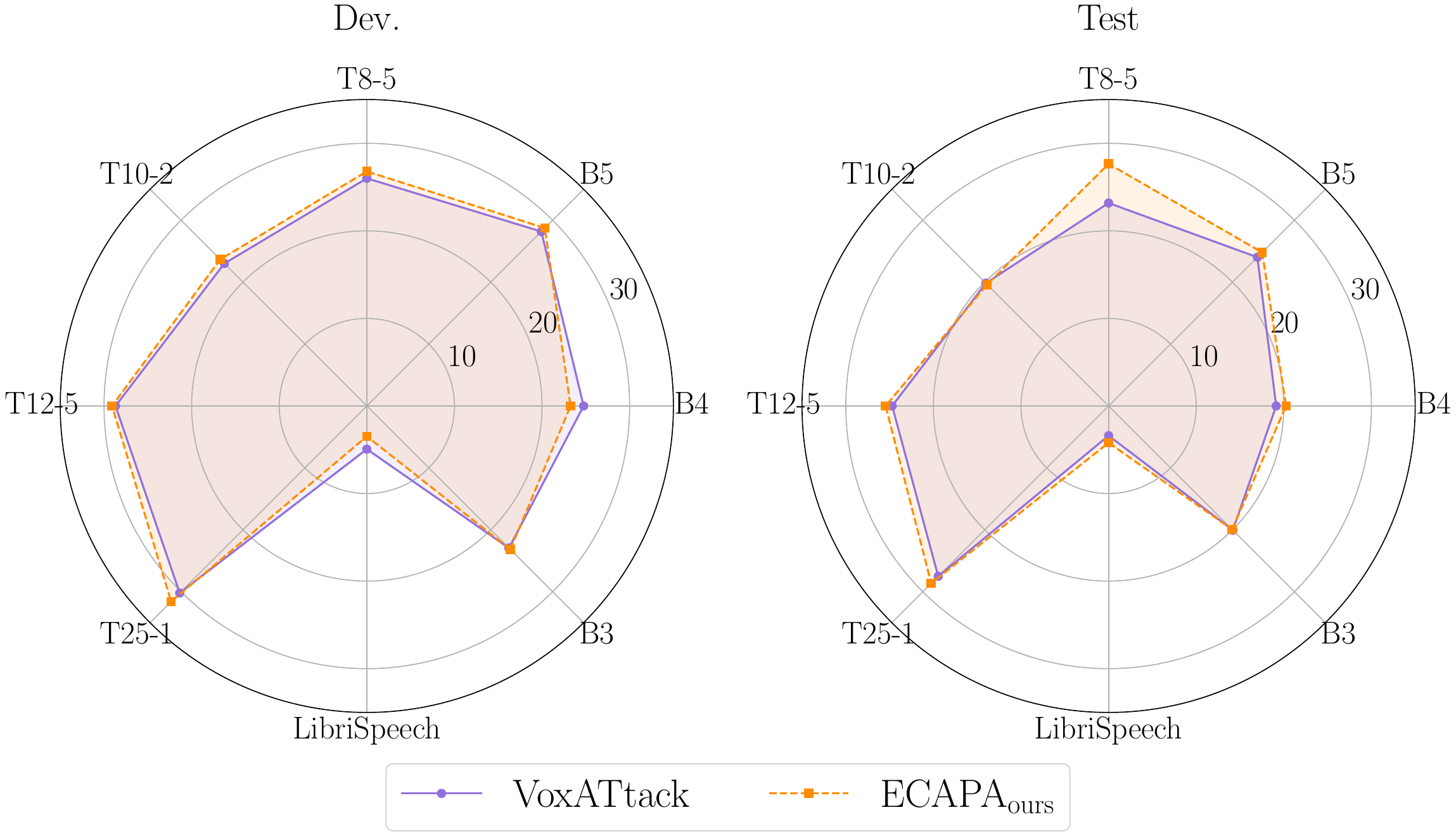}
    \caption{Performance of \proposed against \baselineOurs in \ac{EER}~$[\%]$~\mbox{$(\downarrow)$} across all \anonymization systems and LibriSpeech. The attacker models attack all \anonymization systems simultaneously}
    \label{fig:modality_comparison}
\end{figure}

\subsection{Cross-system evaluation}

\label{sec:gen_attacks}
Next, we evaluate our attacker's robustness when presented with all seven \anonymization systems simultaneously along with the original LibriSpeech dataset. A key question is whether attacks that succeed against individual \anonymization methods can maintain their effectiveness without prior knowledge of which specific model was applied. Previous research suggests this is rather challenging. In particular, the authors in \cite{fine_tune_titanet} demonstrated that TitaNet-Large performs at near-random chance levels against all systems when trained to counter all \anonymization schemes concurrently.

\Cref{tab:gen_attacks} details the scores of our model when trained with all \anonymization systems. 
Contrary to previous findings, our attacks maintained \acp{EER} comparable to those observed in the single-system scenarios, with minimal performance degradation and, in some cases, slight improvements.
Furthermore, our model performed well on the original data, i.e., as a standard \ac{ASV}, relative to the baseline, which was exclusively trained on LibriSpeech.

The largest downgrade was observed with system B4, showing only a $1.3\%$ increase in \ac{EER}. Notably, system T10-2 presented an intriguing anomaly: Despite previously providing the second strongest protection among all systems, it demonstrated the weakest performance in our cross-system setting with an $\textrm{EER}_{\textrm{avg}}$ reduction of $7.6\%$. We examine this unexpected vulnerability further in \cref{sec:augmentation}.

Finally, to inspect whether text fusion alone could explain the discrepancy with \cite{fine_tune_titanet}, we analyzed the performance of \baselineOurs in the same settings. As shown in \Cref{fig:modality_comparison}, \baselineOurs achieved performance nearly identical to \proposed across all systems, with T8-5 being the only exception. These results show that the model has, in fact, improved relative to the single-system case in \cref{sec:main_results}. This directly contradicts previous findings and warrants future investigation.

\begin{figure}[t]
    \centering
    \includegraphics[width=\columnwidth]{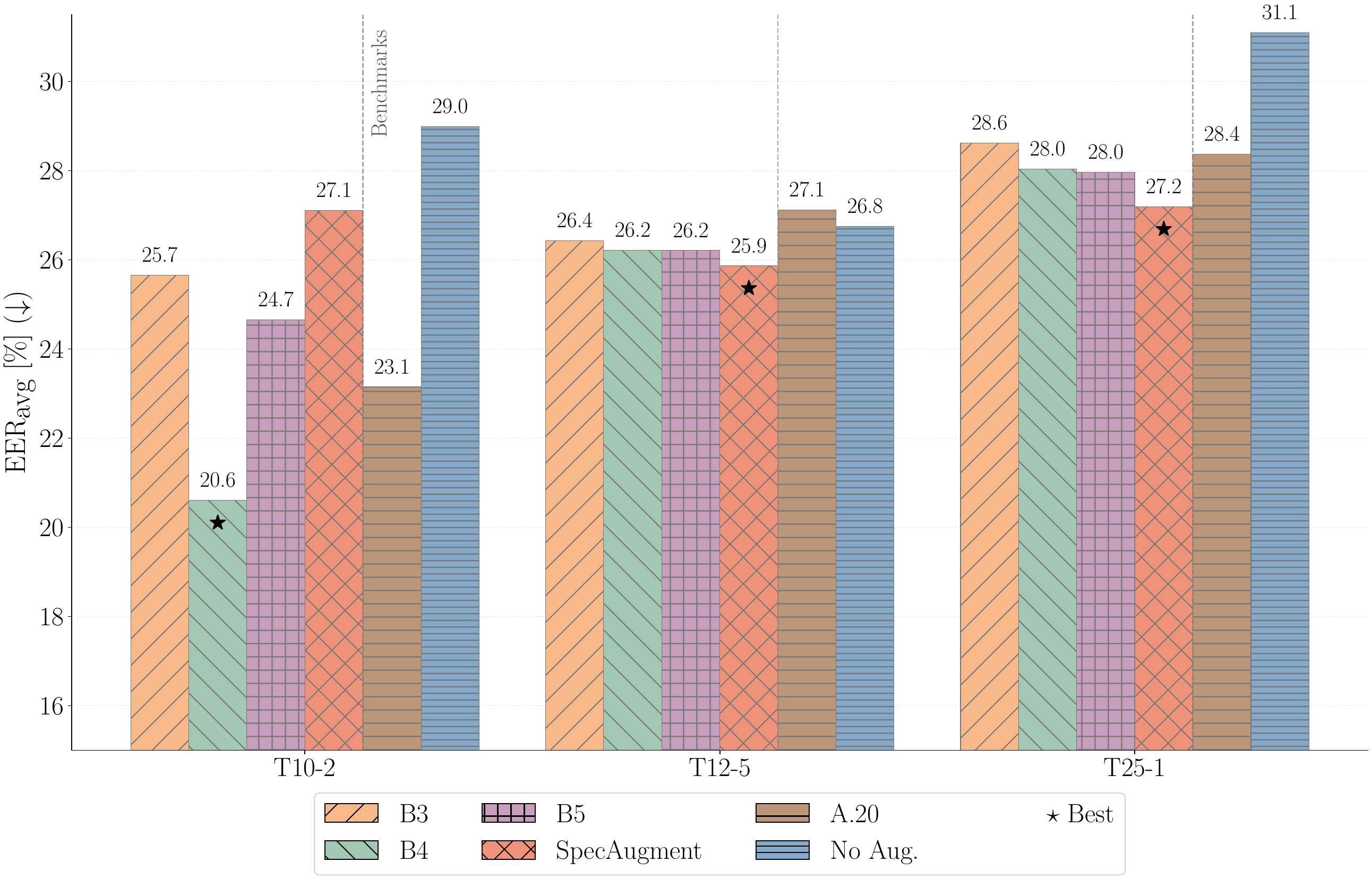}
    \caption{Effect of different augmentations on \proposed against T10-2, T12-5, and T25-1 reported in $\textrm{EER}_{\textrm{avg}}$ $[\%]$~\mbox{$(\downarrow)$}}
    \label{fig:augmentation}
\end{figure}

\subsection{Utilizing \anonymized speech for data augmentation}
\label{sec:augmentation}
Building on our findings in \cref{sec:gen_attacks}, we explore the potential of incorporating \anonymized data into training as an effective augmentation strategy.
To achieve this, we augmented the three most robust anonymizers (ranked by $\textrm{EER}_{\textrm{avg}}$) T10‑2, T12‑5, and T25‑1 with \anonymized speech generated by B3, B4, or B5.
For comparison, we used SpecAugment \cite{park19e_interspeech} as a data augmentation benchmark.

As illustrated in \Cref{fig:augmentation}, our augmented models outperformed A.20 by $2.5\%$ on T10-2 and by $1.2\%$ on T25-1, effectively closing performance gaps where our approach previously underperformed. Significant improvements are also observed when compared to our non-augmented models.
Augmentations enhanced our attacks across all configurations, irrespective of the method or target \anonymization system.
The improvement magnitude, however, varied substantially depending on the evaluated system. Anonymized data provide comparable benefits to SpecAugment on T12-5 and T25-1, but showed an exceptional $6.5\%$ improvement on T10-2.

For T12-5, the best augmentation strategy yielded only modest gains under $1\%$. Augmentations showed larger advantages against T25-1, with SpecAugment reducing $\textrm{EER}_{\textrm{avg}}$ by $3.9\%$. System T10-2 was particularly vulnerable to all augmentation schemes, with a striking $8.4\%$ drop in $\textrm{EER}_{\textrm{avg}}$ when augmented with B4. While this drop could be partly attributed to the use of a neural audio-codec in both systems, this factor alone does not explain why all augmentation methods were consistently more effective against T10-2 than against T12-5 or T25-1. 
Further analysis showed that training \proposed and \baselineOurs on T10‑2 results in the fastest convergence relative to all other anonymization systems, which may explain the increased effectiveness of augmentations on this particular system.

Beyond T10-2's specific vulnerabilities, these findings expose a significant privacy risk: Attackers can leverage known \anonymization methods to disproportionately improve their attacks against unknown systems. This finding opens promising avenues for future research in both offensive and defensive speech \anonymization strategies.

\section{Conclusion}
\label{sec:conclusion}

This work presented \proposed, a multimodal attacker model that exploits both acoustic and textual information to compromise voice \anonymization systems. We adopted a dual-branch architecture combining ECAPA-TDNN for audio embeddings and BERT for textual features, followed by confidence-based fusion. 
Our experiments demonstrate that incorporating textual information enhances attacks against \anonymization systems, with \proposed outperforming the top-ranking \ac{VPAC} systems on five out of seven benchmarks.
By applying data augmentation using \anonymized speech and SpecAugment, \proposed achieves state-of-the-art performance across all seven \ac{VPAC} benchmarks.
Our analysis reveals a potential 
vulnerability in current \anonymization datasets, where speaker-specific linguistic patterns can be exploited by attackers. Furthermore, we have shown that augmenting with \anonymized speech can serve as a successful strategy against certain \anonymization models. Finally, contrary to previous findings, our results indicate that a single attacker system can effectively target multiple \anonymization models simultaneously.

\section*{Acknowledgment}
\label{sec:ack}
Part of this work, including sentence reformulation, code support, and Figures 2-3, was generated or assisted by Claude Sonnet 3.7 thinking and GPT-o1 models. Content was reviewed or edited by the authors.

\clearpage
\IEEEtriggeratref{16}
\bibliographystyle{IEEEtran}
\bibliography{utils/refs25}

\end{document}